\documentclass[letter,prl,showpacs,twocolumn]{revtex4-1}
\usepackage{graphicx}
\usepackage{dcolumn}
\usepackage{bm}
\usepackage{amssymb}
\usepackage{amsmath}
\usepackage{epsfig}
\usepackage{times}

\input{tcilatex}

\begin{document}

\title{Quantum statistical imaging of particles without restriction of the
diffraction limit}
\author{Jin-Ming Cui}
\author{Fang-Wen Sun}
\email{fwsun@ustc.edu.cn}
\author{Xiang-Dong Chen}
\author{Zhao-Jun Gong}
\author{Guang-Can Guo}
\affiliation{Key Lab of Quantum Information, University of Science and Technology of
China, Hefei 230026 }
\date{\today }

\begin{abstract}
A quantum measurement method based on the quantum nature of anti-bunching
photon emission has been developed to detect single particles without the
restriction of the diffraction limit. By simultaneously counting the
single-photon and two-photon signals with fluorescence microscopy, the
images of nearby Nitrogen-Vacancy centers in diamond at a distance of $%
8.5\pm 2.4$ \textrm{nm} have been successfully reconstructed. Also their
axes information was optically obtained. This quantum statistical imaging
technique, with a simple experimental setup, can also be easily generalized
in the measuring and distinguishing of other physical properties with any
overlapping, which shows high potential in future image and study of coupled
quantum systems for quantum information techniques.
\end{abstract}

\maketitle

\affiliation{Key Lab of Quantum Information, University of Science and Technology of
China, Hefei 230026 }

The measurement of physical quantities is not only a major goal but also an
active impulsion for scientific research. Especially, the imaging of nearby
particles is important for modern science \cite{Alivisatos,Patterson}. The
precision with which two nearby particles can be resolved is classically
restricted by the optical diffraction limit. Imaging methods that used
distinguishing information based on the photons emitted from different
particles have been proposed to achieve precision beyond the diffraction
limit \cite{Ash,Denk,Klar,Rust,Betzig,Hess,Dertinger}. When the emitted
photons have the same properties, distinguishing nearby particles which are
separated by distances much less than the diffraction limit is difficult.

Recently, phenomena from quantum mechanics have been used to improve the
measurement and applied to some special purposes which cannot be performed
by classical method. Such quantum techniques are being applied to enhance
the precision of measurements beyond the classical limit \cite%
{Giovannetti,LIGO}. However, many quantum-based protocols to improve the
measurement used the quantum entanglement. They are fragile because of
quantum decoherence \cite{Nagata,Sun,Xiang}. For practical purpose, stable
quantum phenomena should be applied in the measurement. There are proposals
to enhance the imaging resolution based on the quantum statistics \cite%
{Dertinger,Schwartz}. Here, we have developed a quantum statistical imaging
(QSI) method to detect single particles without the restriction of the
diffraction limit. In the quantum regime, the situation for particles
imaging is different because each particle emits only one photon and shows
the single-photon antibunching effect \cite{Kimble}. By detecting the photon
coincident counts, the particles can be imaged and resolved even when they
are almost completely overlapping and the emitted photons are identical.
Here, Nitrogen-vacancy center (NVC) in diamond was used in this experimental
demonstration. Single NVC has shown its good quality as a single-photon
source \cite{Kurtsiefer,Babinec}. When two NVCs are close to each other,
i.e., within tens of nanometers, the strong dipole-dipole interaction can be
applied in quantum information techniques \cite{Maurer,Neumann}. Diamond
nano-crystals with NVCs have been successfully used to image biological
processes \cite{Chang,McGuinness}. The ability to image and distinguish two
nearby NVCs is becoming increasingly important in physics \cite%
{Maurer,Neumann,Rittweger} and biology \cite{Chang,McGuinness}.

NVCs are usually detected by scanning confocal optical fluorescence
microscopy \cite{Gruber}, in which the single-photon intensity of
spontaneous emission is measured to characterize the optical images of the
centers. The spontaneous emission from the NVCs, which were fabricated by
nitrogen ions implantation, is collected into a single-mode fiber. The
collected photons are then split into two paths by a fiber beam splitter
with a transmissivity of T and a reflectivity of R ($T+R=1$), to form a
Hanbury-Brown-Twiss interferometer \cite{HBT}. Finally, the separated beams
are detected with two single-photon detectors. When two NVCs ($A$ and $B$)
are imaged, the single-photon intensity at position ($x$, $y$) from the
single-photon detectors $D_{1}$ and $D_{2}$ can be expressed as

\begin{eqnarray*}
\left\langle I_{1}^{D1}(x,y)\right\rangle &=&T[\left\langle
I_{A}(x,y)\right\rangle +\left\langle I_{B}(x,y)\right\rangle ]\text{,} \\
\left\langle I_{1}^{D2}(x,y)\right\rangle &=&R[\left\langle
I_{A}(x,y)\right\rangle +\left\langle I_{B}(x,y)\right\rangle ]\text{,}
\end{eqnarray*}%
where $\left\langle I_{A}(x,y)\right\rangle $ and $\left\langle
I_{B}(x,y)\right\rangle $ are the single-photon rates from NVCs $A$ and $B$,
respectively. Therefore, the single photons from the two NVCs are

\begin{eqnarray}
\left\langle I_{1}(x,y)\right\rangle &=&\left\langle
I_{1}^{D1}(x,y)\right\rangle +\left\langle I_{1}^{D2}(x,y)\right\rangle
\notag \\
&=&\left\langle I_{A}(x,y)\right\rangle +\left\langle
I_{B}(x,y)\right\rangle \text{.}  \label{I1}
\end{eqnarray}%
When the distance between the two NVCs is within the optical diffraction
limit, $\left\langle I_{A}(x,y)\right\rangle $ and $\left\langle
I_{B}(x,y)\right\rangle $ are overlapping and hardly distinguishable from
the single-photon intensity $\left\langle I_{1}(x,y)\right\rangle $. If
two-photon coincident counts are measured, then the two photons must come
from two NVCs and never from the same NVC because a single NVC only emits
one photon. This attribute demonstrates a genuine quantum characteristic,
namely, the photon anti-bunching effect with $\left\langle
:[I_{A}(x,y)]^{2}:\right\rangle =\left\langle
:[I_{B}(x,y)]^{2}:\right\rangle =0$, where $::$ represents normal ordering
\cite{QO}. Therefore, the two-photon intensity will be

\begin{equation}
\left\langle I_{2}(x,y)\right\rangle =\eta _{2}(1+K)RT\left\langle
I_{A}(x,y)\right\rangle \left\langle I_{B}(x,y)\right\rangle \text{,}
\label{I2}
\end{equation}%
where $\eta _{2}$ is the two-photon detection constant, which is based on
the imperfections from the photon collection efficiency, path loss,
detection efficiency of the single-photon detectors and coincident detection
windows in the experiment. $K$ describes the quantum
indistinguishability-induced bunching effect of two photons \cite%
{Santori,Sun09}. Usually, $K=0$ in many measurements without special
spectral filtering. For example, the photon from NVC has very broad phonon
bandwidth comparing to its narrow zero-phonon line width \cite{Gruber},
leading $K=0$ in the present measurement. Therefore, simply from $%
\left\langle I_{1}(x,y)\right\rangle $ and $\left\langle
I_{2}(x,y)\right\rangle $, the values of $\left\langle
I_{A}(x,y)\right\rangle $ and $\left\langle I_{B}(x,y)\right\rangle $ can be
obtained, even if they are completely overlapping. Subsequently, the optical
images of the two NVCs can be reconstructed and distinguished. For $N$
particles, $m$-th ($1\leqslant m\leqslant N$) order coincidence measurement
will give
\begin{equation*}
\left\langle I_{m}(x,y)\right\rangle =\eta _{m}\dsum [\left\langle
I_{A}(x,y)\right\rangle \left\langle I_{B}(x,y)\right\rangle ...]_{m\text{
different points}}\text{,}
\end{equation*}%
where $\eta _{m}$ is the $m$-photon detection constant and the photon
indistinguishability induced photon bunching effect is also neglected. There
are $N$ independent values, where images of each particles can be solved and
reconstructed. There is no need to have any assumption on the distribution
function of the image or the point spread function \cite{Schwartz}. This
technique utilizes a genuine quantum phenomenon to produce this result
without classical parallelism.
\begin{figure}[tbp]
\begin{centering}
\includegraphics[width=7.5cm]{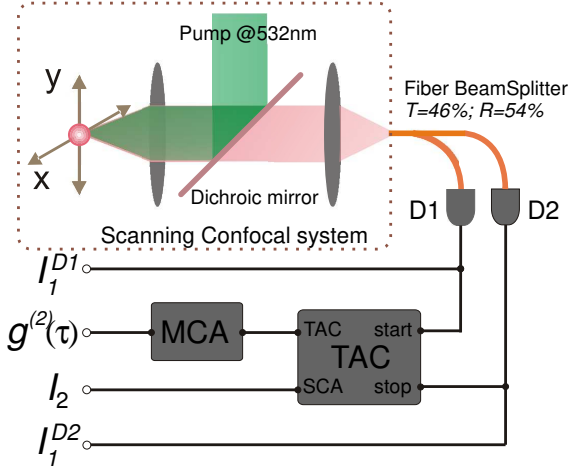}
\par\end{centering}
\par
\centering{}
\caption{(color online) Schematic of measurement setup. Confocal NV
fluorescent photons are splitted by a fiber beam splitter and sent to single
photon detectors $D1$ and $D2$. TAC is used to get coincidence counts and
MCA is to implement autocorrelation measurement ($g_{c}^{(2)}(\protect\tau )$%
). $I_{1}^{D1}$, $I_{1}^{D1}$, and $I_{2}$ are $D1$, $D2$ and coincidence
count rates, respectively. }
\label{fig:Schem}
\end{figure}

In the experiment, NVCs are excited by a continuous $532$ \textrm{nm} green
laser, and the experimental setup is shown in Fig. \ref{fig:Schem}. To
record the two-photon counting rate $\left\langle I_{2}\right\rangle $ at
the same time, a single-channel analyzer (SCA) on a time-amplitude converter
(TAC) is used to obtain the coincidence counts of $D1$ and $D2$, with a SCA
window width of $t_{w}=2$ \textrm{ns} at $\tau =0$ because a single NV
center has an anti-bunching decay with a width of approximately $20$ \textrm{%
ns}. The two-photon autocorrelation measurement ($g_{c}^{(2)}(\tau )$) is
performed by the multi-channel analyzer (MCA). Using scanning confocal
microscopy, $\left\langle I_{1}\right\rangle $ and $\left\langle
I_{2}\right\rangle $ for each position are recorded to construct
single-photon and two-photon images. The coincidence measurement is
conducted in start-stop mode in the TAC, and
\begin{equation*}
\eta _{2}=\frac{2t_{w}\left\langle I_{1}^{D1}\right\rangle \left\langle
I_{1}^{D2}\right\rangle }{RT(\left\langle I_{1}^{D1}\right\rangle
+\left\langle I_{1}^{D2}\right\rangle )^{2}}\text{,}
\end{equation*}%
with $R=54\%$ and $T=46\%$.
\begin{figure}[tbp]
\begin{centering}
\includegraphics[width=7.5cm]{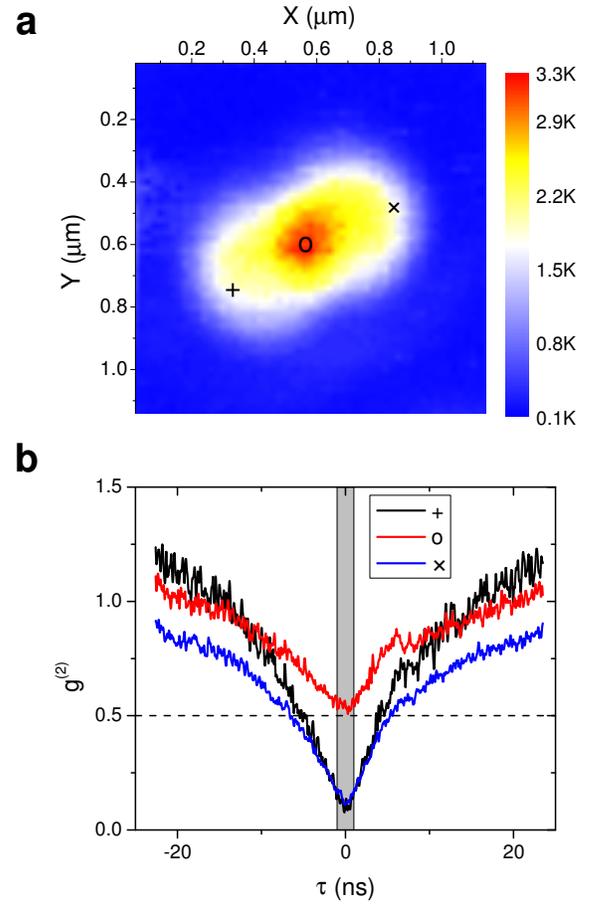}
\end{centering}
\caption{(color online) (a) Confocal image of two NVCs apart from $366.1$%
\thinspace \textrm{nm}. (b) $g_{c}^{(2)}(\protect\tau )$ at different points
in (a). The gray window indicates the SCA window $t_{w}=2\mathrm{ns}$ to
measure $I_{2}$. }
\label{fig:g2}
\end{figure}
\

Two nearby two-NVC pairs were measured. Fig. \ref{fig:g2} (a) displays a
confocal scanning image of the first two-NVC pair. Fig. \ref{fig:g2} (b)
displays $g_{c}^{(2)}(\tau )$ with background noise subtraction. The
position with maximal intensity marked as \textquotedblleft $\circ $%
\textquotedblright\ is at the overlapping area of two NVCs, where both NVCs
are excited in the pump focus with $g_{c}^{(2)}(0)$ is about $0.5$. However,
while the confocal spot is on side position marked as \textquotedblleft $+$%
\textquotedblright\ or \textquotedblleft $\times $\textquotedblright , only
one NVC are effectively excited as another NVC is out of exciton spot.
Therefore, $g_{c}^{(2)}(0)$ is close to zero as a single NVC. From marked
positions \textquotedblleft $+$\textquotedblright\ to \textquotedblleft $%
\circ $\textquotedblright\ in Fig. \ref{fig:g2} (a), it is clear to see the
variation from single to double NVCs.
\begin{figure}[tbp]
\begin{centering}
\includegraphics[width=7.5cm]{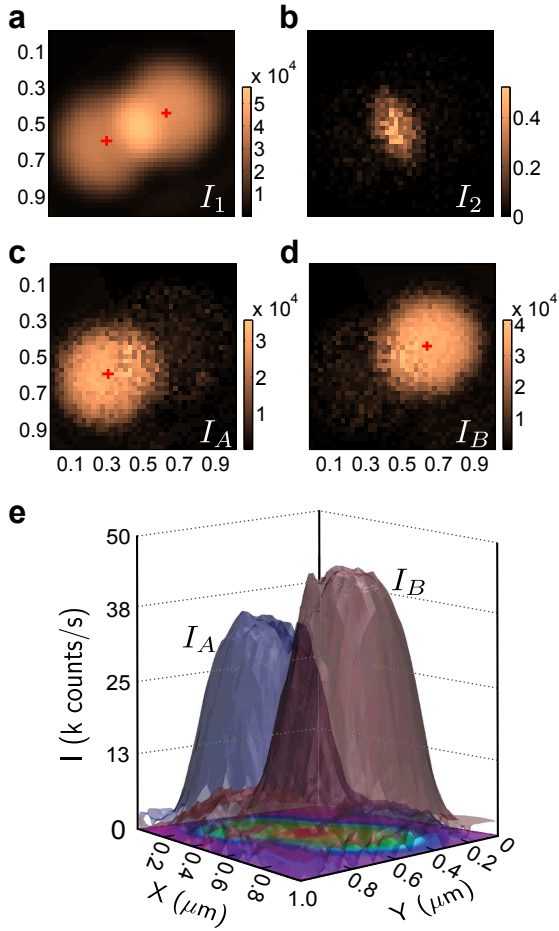}
\par\end{centering}
\caption{(color online) Optical images of two single NVCs at a distance of $%
366.1\pm 2.8$ \textrm{nm}. (a) and (b) show $I_{1}$ and $I_{2}$ for the
single-photon and two-photon counts. (c) and (d) are images of each NVC ($%
I_{A}$, $I_{B}$), respectively. The red crosses mark the positions of the
NVCs with an uncertainty of $1/20$ the length of each cross. The positions
were obtained using a two-dimensional (2D) Gaussian fitting of $I_{A}$ or $%
I_{B}$. (e) A 3D image of the two NVCs.}
\label{fig:Far}
\end{figure}

For the first pair, the single-photon intensity (($\left\langle
I_{1}(x,y)\right\rangle $) and two-photon intensity ($\left\langle
I_{2}(x,y)\right\rangle $) were recorded at the same time (Fig. \ref{fig:Far}
(a, b)). In Fig. \ref{fig:Far} (b), the image of the two-photon intensity
has a narrower width in the overlapping region of the two NVCs. However,
these images do not provide spatially resolved images of the two NV centers.
The image of a single NVC should have a single peak and a continuous
envelope. Using $\left\langle I_{1}(x,y)\right\rangle $ and $\left\langle
I_{2}(x,y)\right\rangle $, the photon intensities $\left\langle
I_{A}(x,y)\right\rangle $ and $\left\langle I_{B}(x,y)\right\rangle $ of the
two particles can be obtained by solving Eq. (\ref{I1}) and Eq. (\ref{I2}).
The images of the two particles, $I_{A}$ and $I_{B}$, were reconstructed and
are shown in Fig. \ref{fig:Far}(c, d). By fitting the data, the distance
between the NVCs was determined to be $366.1\pm 2.8$ \textrm{nm}, which is
at the edge of the diffraction limit. Fig. \ref{fig:Far} (e) shows a
three-dimensional (3D) image of the two nearby NVCs, and the overlapping can
be observed.
\begin{figure}[tbp]
\begin{centering}
\includegraphics[width=7.5cm]{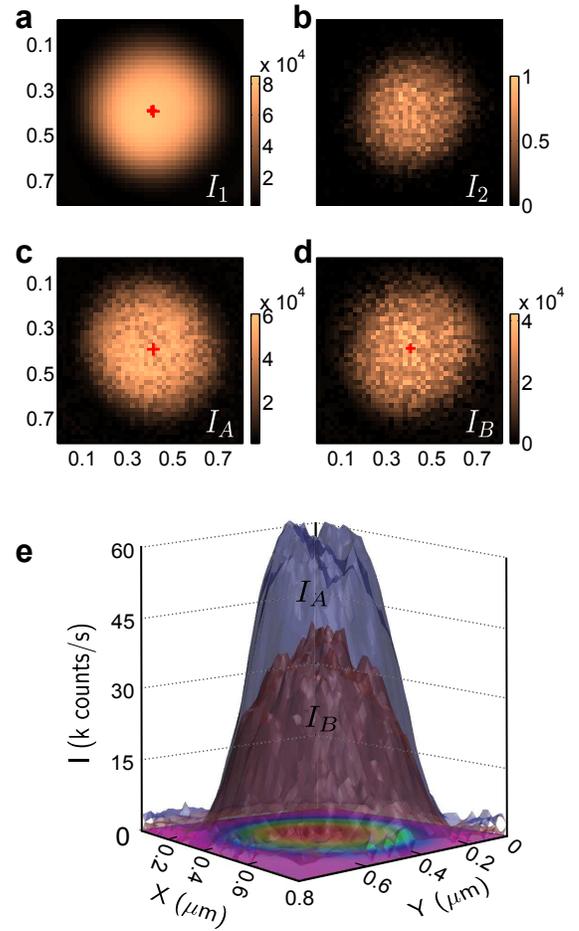}
\par\end{centering}
\caption{(color online) Optical imaging of two single NVCs with a small
separation. The images are organized as in Fig.\protect\ref{fig:Far}. The
distance between the NVCs was determined to be $8.5\pm 2.4$ \textrm{nm} by
fitting $I_{A}$ and $I_{B}$.}
\label{fig:Near}
\end{figure}

Another pair of NVCs with a much smaller separation was also measured and
distinguished. The confocal scanning image, the result of photon
autocorrelation measurement and the spectrum show that there are two NVCs
(see Supplementary Figure S2). The single-photon and two-photon images are
shown in Fig. \ref{fig:Near} (a, b). From the single-photon intensity, the
two NVCs are well overlapping and cannot be distinguished. Using the QSI
method, their images, $I_{A}$ and $I_{B}$, can be obtained, shown in Fig. %
\ref{fig:Near} (c, d). The distance between the centers was determined to be
$8.5\pm 2.4$ \textrm{nm} by fitting $I_{A}$ and $I_{B}$. This distance is
much smaller than the diffraction limit, and the NVCs cannot be readily
distinguished by the classical method \cite{Maurer}. Here, the resolution
(see Supplementary for details) was determined by the number of recorded
photons and the experimental setup. In the present measurement, the error is
about $0.9$ \textrm{nm} with $10^{5}$ coincident photon counts and the setup
repeat resolution for $x$ or $y$ axis is about $0.7$ \textrm{nm}.

\begin{figure}[tbp]
\begin{centering}
\includegraphics[width=7.5cm]{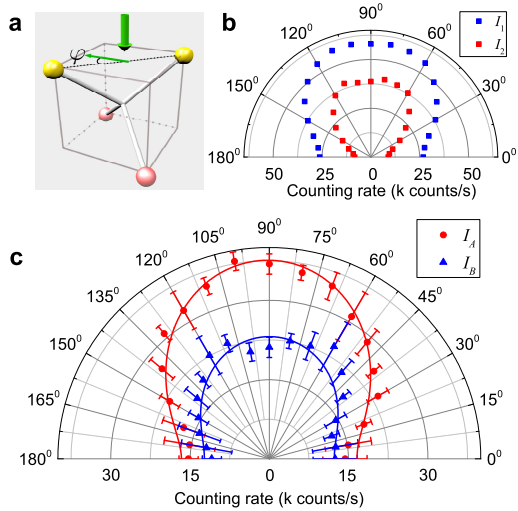}
\par\end{centering}
\caption{(color online) (a) The schematic of two sets of axes for
[100]-oriented NVC sample according to the polarization of pump beam. They
are in the plane of $\protect\varphi =0^{\circ }$ or $\protect\varphi %
=90^{\circ }$. (b) Single-photon and two-photon emission intensity versus
the polarization angle of pump beam. The two-photon intensity was amplified
by $5\times 10^{3}$ for view. (c) Separated intensities of two NVCs versus
the polarization angle of pump beam.}
\label{fig:axes}
\end{figure}

Besides imaging nearby particles, such a quantum measurement method can be
easily generalized to measure and distinguish other properties even with
high overlapping. For example, the axes of the NVCs can be measured. The
spontaneous emission rates vary with the polarization of the pump beam
according to different axes of NVCs \cite{Epstein,Alegre}. With polarized
optical pump for our [100]-oriented sample, number of possible orientations
of a given center is reduced from four to two, which are in the plane of $%
\varphi =0^{\circ }$ or $\varphi =90^{\circ }$ as shown in Fig. \ref%
{fig:axes} (a). Here with QSI, the axes of the pair of NVCs at the distance
of $8.5$ \textrm{nm} have been obtained. With different polarized pump beam,
the single-photon and two-photon counts are shown in Fig. \ref{fig:axes}
(b). Simply, the emission intensity of each NVCs with the angle of
polarization can be obtained in Fig. \ref{fig:axes} (c). Correspondingly,
the axes of the two NVCs are same in the plane of $\varphi =0^{\circ }$. The
data can be fitted with $I_{A(B)}=\alpha _{A(B)}+\beta _{A(B)}\cos
^{2}\varphi $ \cite{Alegre} with $\alpha _{A(B)}=37.5\pm 0.4(23.1\pm 0.5)$
and $\beta _{A(B)}=-21.0\pm 0.8(-10.8\pm 0.9)$. The small dip at $\varphi
=90^{\circ }$ for NV$_{B}$ may come from its depth in the diamond \cite%
{Epstein}.

In addition to the demonstrated symmetric envelopes of
the photons from NVCs, the QSI method can be applied to detect and distinguish images with other
continuous envelopes. Furthermore, the QSI method can be used to detect other particles and
distinguish other degrees of freedom, such as lifetime,
frequency, and polarization. Because the photons from each particle
can be recorded simultaneously and distinguished without decoupling nearby
particles, the QSI method can be used to detect their separate
dynamics, as well as the coupling between them. Also, the experimental setup
for QSI is simple. It does not need complicated pump beams \cite%
{Maurer,Rittweger} or the assistance of other control systems \cite%
{Maurer,Neumann}. In the current experiment, the single-photon counts were
about $20\mathrm{K/s}$ and the total detection efficiency was about $0.15\%$,
including the loss at the collection objective ($97\%$), pass loss ($90\%$)
and the loss at the detector ($50\%$). However, in principle, all of these
performances can be much improved. The collection efficiency can be enhanced
by a factor of $8$ with structured interface \cite{Marseglia}. The loss from
beamsplitter can be removed with additional detectors \cite{Xiang}. The loss
at the detector can be overcome with high quantum efficiency. If
pulsed laser is used, there is no need the $2\mathrm{ns}$ gate and all the
photons within its lifetime can be collected, which is about $6$-fold
enhancement in the two-photon counts. Along with other low loss filters, the
total two-photon detection rate can be improved by over three orders of
magnitude. For the image of NVCs, the sample was moved at $40\times 40$
locations (pixels) and photon counts were collected at each location. A
gated CCD can be used to give additional enhancement in the data
collection. With above improved techniques, the total data collection time
will be less than one minute for the two NVCs with same resolution, compared
to current $50$ hours. It is very promising in the scalable application for
multi-partite measurement and the detection of particles which emit small numbers of photons because of bleaching or blinking.
In those cases, the resolution is limited by the
total photons. However, it will still have very good
resolution. For example, $10^{5}$ photons can offer
the sub-nanometer resolution.

In summary, a QSI method was demonstrated based on the unique quantum
behavior of anti-bunching emission photons. Two well-overlapping NVCs can be
spatially resolved. Also, the axes of the two NVCs are measured even their
distance is within $8.5$ \textrm{nm}. With high order of coincident
multi-photon measurements, additional single NVCs can be imaged and
distinguished. The scalable QSI with high order coincidence counts will be
applied in the multipartite interaction for quantum information techniques
and modern physics.

\begin{acknowledgments}
This work was supported by the 973 Programs (No.2011CB921200 and No.
2011CBA00200), the National Natural Science Foundation of China (NSFC) (No.
11004184), the Knowledge Innovation Project of the Chinese Academy of
Sciences (CAS), and the Fundamental Research Funds for the Central
Universities.
\end{acknowledgments}

\end{document}